\documentstyle[pre,aps,epsf]{revtex}
\newcommand{\be}{\begin{equation}}
\newcommand{\ee}{\end{equation}}
\newcommand{\beq}{\begin{eqnarray}}
\newcommand{\eeq}{\end{eqnarray}}
\def\({\left(}
\def\){\right)}
\newcommand{\non}{\nonumber\\}
\def\Ai{{\rm Ai}}

\begin{document}
\title{Ballistic aggregation: a solvable model of irreversible many particles
dynamics}
\author{L. Frachebourg\footnote{supported by
the Swiss National Foundation for Scientific Research.}
and Ph. A. Martin}
\address{Institut de Physique Th\'eorique}
\address{Ecole Polytechnique F\'ed\'erale de Lausanne}
\address{CH-1015 Lausanne, Switzerland}
\author{J. Piasecki}
\address{Institute of Theoretical Physics}
\address{University of Warsaw}
\address{00 681 Warsaw, Poland}

\date{\today}

\maketitle

\begin{abstract}

The adhesive dynamics of a one-dimensional aggregating gas of point
particles is rigorously described. The infinite hierarchy of kinetic 
equations for the distributions of clusters of nearest neighbours
is shown to be equivalent to a system of two coupled equations for a large
class of initial conditions. The solution to these nonlinear equations is
found by a direct construction of the relevant probability distributions 
in the limit of a continuous initial mass distribution.
We show that those limiting distributions are identical to those 
of the statistics of shocks in the Burgers turbulence. The analysis relies on
a mapping on a Brownian motion problem with parabolic constraints.

\end{abstract}

\vskip 1truecm

\section{Introduction}

Rigorous solutions of the many-body dynamics are extremely rare. The
present paper yields up such a precious case, reviewing parts of the
existing works and completing the research
which has been carried out in recent years.

The system under study is a one-dimensional gas of point particles  
forming aggregates through perfectly inelastic, adhesive collisions. The
motion between collisions being free, the process is called
ballistic aggregation. This dynamics is deterministic, randomness occurs
only through the distribution of initial datas. In the context of statistical
mechanics, the model was introduced and studied numerically in \cite{pom90}
and further investigated analytically in \cite{jpias92}, \cite{Mar-Pia} 
and \cite{frach99}. However in the context of fluid dynamics, 
it has been recognized
much earlier (see e.g. \cite{burgers}, \cite{kida}) that the 
evolution of shocks in 
the inviscid limit of the (decaying) one-dimensional Burgers 
equation obeys the laws of ballistic aggregation.
Moreover when the initial particle velocities are Gaussian and independent 
(equivalently when the initial Burgers field is a white noise) both models
are isomorphic to a problem of Brownian motion under parabolic constraints.
The latter problem has eventually received a solution in closed 
analytical form
\cite{frach99}, \cite{FraMar99} which enables to predict the exact 
distribution of shocks in the decaying Burgers turbulence, and correspondingly
the distribution of masses and velocities of the aggregating particles.
These close connections between problems of different origins is 
an attractive feature
of ballistic aggregation.
The purpose of this work is to establish precisely the above 
mentioned relations
and to give an exact and fairly complete description of ballistic aggregation.

There are essentially two approaches. The first one 
consists in writing the dynamical hierarchy of equations coupling
the many-particle distribution functions using the tools of kinetic theory, 
and hopefully solving these equations.
The second route is by explicitly constructing the dynamics from suitably
chosen initial conditions: the construction should of course yield a solution 
of the kinetic equations. Both approaches are discussed here and both will make
the link with the dynamics of Burgers shock waves manifest. 

After recalling the way in which one can describe the state of an infinite gas
(Section \ref{sec2}), we present in Section \ref{sec3} 
the derivation of the infinite 
hierarchy of kinetic equations coupling the time evolution of 
distributions of many-particle clusters of nearest neighbours. The derived
hierarchy, after integration over momenta of the aggregates, turns out to
be identical to the hierarchy found in the study of the Burgers
turbulence \cite{kida}. 

In Section \ref{sec4} an original result is derived. 
The infinite dynamical hierarchy 
is shown to be compatible with the factorization of the many-body
distributions into products of two-particle conditional distributions and
the one-particle density. Such a factorization is propagated by the
ballistic aggregation. This yields an exact closure of the hierarchy 
and permits to reduce 
it to a system of two coupled equations (Section \ref{sec5})
which allow self-similar
solutions (Section \ref{sec6}). 

The Section \ref{sec7} is devoted to an explicit construction of 
the particle distributions.
The construction, which has already been presented in \cite{Mar-Pia}, 
uses a simple initial condition
which enables to reformulate the problem within the theory of Brownian motion.
To establish the relation with shock wave dynamics at this level, 
one must envisage the 
delicate question of the continuum limit.
Indeed the Burgers velocity field takes values in the continuum, whereas masses
of aggregates, resulting of sums of elementary initial masses, have discrete 
values. A strict isomorphism will only be obtained in the continuum 
limit of the
aggregation process letting the distribution of initial masses tend 
to a continuous and uniform
mass density. In terms of Brownian motion this leads to the subtle problem
of controlling the cumulated effect of Brownian excursions in 
infinitely many small 
time intervals. This problem is addressed and extensively 
discussed in Section \ref{sec8}:
in the continuum limit the particle distributions become 
identical to those of the shocks in Burgers dynamics
so that all the results of \cite{FraMar99} immediately 
apply. All distributions of
order three and more factorize in the way described in Section \ref{sec4}, 
so giving
a complete statistical description of the aggregation process. 
In Section \ref{sec9} 
the exact predictions for the mass and velocity distributions 
are compared to existing bounds 
as well as to the findings of previously formulated approximate theories.
In particular significant discrepancies with mean field theory 
appear in the domains
of large and small masses.
The asymptotic form of the two-particle correlation function at large distance
displays the range of spatial correlations induced by the aggregation process:
one finds a very rapid decay of such correlations. 
In the last section, we briefly indicate by which mechanisms 
our explicit solution
verifies the closed dynamical equation derived in Section \ref{sec4}
(and hence the whole hierarchy),
thus completing our analysis of the model. 

\section{States of the aggregating gas}\label{sec2}

At any time $t>0$, the system is composed of point particles moving in
$R^{1}$. The state of a given particle is entirely characterized by 
specifying its position $X$, its momentum $P$, and its mass $M$. 

Consider a closed space interval $[L_{1},L_{2}]$ , where $L_{1}<L_{2}$. We 
denote by
\be
\mu_{k}(1,2,\ldots ,k;t|L_{1},L_{2}) \label{1}
\ee
the probability density for finding precisely $k$ particles
within the interval $[L_{1},L_{2}]$ at time $t$, occupying the sequence of
states
\be
j \equiv (X_{j},P_{j},M_{j}), \;\;\;\; j=1,2,\ldots ,k \label{2}
\ee
ordered in space according to the inequalities
\be
L_{1} < X_{1} < X_{2} < \ldots  < X_{k} < L_{2}  \label{3} 
\ee
The set of densities $\mu_{k}(1,2,\ldots ,k;t|L_{1},L_{2})$, 
$k=0,1,2,\ldots $, 
defined for arbitrary intervals $[L_{1},L_{2}]$, provides a complete 
statistical description of the state of the infinite volume of the gas.  
Summing up the probability weights for all possible events within 
$[L_{1},L_{2}]$ one gets the normalization condition
\beq
\mu_{0}(t|L_{1},L_{2}) + 
\sum_{k=1}^{\infty}\int d1 \int d2 && \ldots  \int dk\;
\theta (X_{1}-L_{1}) \, \theta (L_{2}-X_{k}) \non
&&\times \prod_{j=2}^{k}\theta (X_{j}-X_{j-1}) 
\mu_{k}(1,2,\ldots ,k;t|L_{1},L_{2})
 = 1
\label{4}
\eeq
where
\be
\int dj \equiv \int_{-\infty}^{+\infty} dX_{j}\int_{-\infty}^{+\infty} dP_{j}
\int_{0}^{+\infty} dM_{j}\; , \label{5}
\ee
and $\theta (X)$ is the unit Heaviside step function.

Knowing the probability densities $\mu_{k}(1,2,\ldots ,k;t|L_{1},L_{2})$ one can
evaluate the densities  $\mu_{k}(1,2,\ldots ,k;t)$ of the nearest neighbours  
configurations  by letting the extremities of the interval $[L_{1},L_{2}]$ 
approach the positions of the first and the last particle, respectively
\be
\mu_{k}(1,2,\ldots ,k;t) \equiv \lim_{L_{1}\to X_{1}}\lim_{L_{2}\to X_{k}}
\mu_{k}(1,2,\ldots ,k;t|L_{1},L_{2})  \label{6}
\ee
In particular, $\mu_{1}(1;t)$ represents simply  the number density of 
particles at time $t$ in the one-particle state $1 \equiv (X_{1},P_{1},M_{1})$.

An alternative way of describing the state of an infinite system consists
in determining the complete set of reduced distributions
$\rho_k(1,2,\ldots ,k;t)$, representing the number density of ordered
k-particle states $(1,2,\ldots ,k)$. 
The difference with respect to the densities
$\mu_{k}(1,2,\ldots ,k;t)$ comes from the fact that the $k$ particles 
need not represent the set of nearest neighbours. Although in evaluating 
$\rho_k(1,2,\ldots ,k;t)$  the inequalities
\be
X_{1}<X_{2}< \ldots  <X_{k} \label{7}
\ee
are still supposed to hold,  one has to consider all configurations of the 
system compatible with the condition that the states $(1,2,\ldots ,k)$ are 
occupied at time $t$, with an  arbitrary number of particles between pairs 
$(j,j+1), \; j=1,\ldots ,k-1$. 
Clearly, there is then no difference at the level of the one-particle 
densities, and, as it has already been mentioned, the equality
\be
\rho_{1}(1;t) = \mu_{1}(1;t)  \label{8} 
\ee
holds. 

For the two-particle reduced density $\rho_{2}(1,2;t)$, the 
corresponding formula reads 
\beq
\rho_{2}(1,2;t) = \mu_{2}(1,2;t) + 
\sum_{r=1}^{\infty}&&\int d1'\int d2' \ldots
\int dr'\, \theta (X_{1}'-X_{1}) \theta (X_{2}-X_{r}')\non 
&& \times \prod_{j=1}^{r-1}\theta (X_{j+1}'-X_{j}') \,
\mu_{2+r} (1,1',2',\ldots ,r',2;t) 
\label{9}
\eeq
The possibility of the presence of $r$ particles $(r=0,1,2,\ldots )$ between
the two particles at $X_{1}$ and at $X_{2}$,  has been taken into 
account in (\ref{9}). By definition (\ref{6}) the density 
$\mu_{2+r}(1,1',2',\ldots ,r',2;t)$ vanishes outside the region 
$ X_{1}<X_{1}'< \ldots  <X_{r}'<X_{2}$, reflecting the linear ordering of the
particles. 

In a similar way (i.e. by considering all possible intermediate states)
one can express $\rho_{k}(1,2,\ldots ,k;t)$ in terms of 
the densities of nearest 
neighbours $\mu_{k+r}(1,2,\ldots ,k+r;t)$,  $r=0,1,2,\ldots$, 
for arbitrary $k=3,4,\ldots$. 
The general relations between both types of description of the states of
an infinite system are discussed in Ruelle's book \cite{ruelle}.

\section{EVOLUTION OF THE DISTRIBUTION OF NEAREST NEIGHBOURS}
\label{sec3}

Our main object here is the study of the dynamics of aggregation. The point 
masses forming the system move freely between collisions. When two of them
collide, they instantaneously merge forming a new particle whose mass 
is equal to the sum of their masses. As the momentum is also conserved,
the aggregate continues the motion along the center of mass trajectory of
the pair it has been formed from. Hence, the microscopic dynamics consists
of periods of a free motion separated by perfectly inelastic (sticky or
adhesive) binary collisions.\\
Suppose that at some moment two particles occupy the states
$a=(X_{a},P_{a},M_{a})$ and $b=(X_{b},P_{b},M_{b})$, with $X_{a}<X_{b}$.
The notation $(a+b)$ will be used  to denote the state 
corresponding to the center of mass motion of such an ordered pair. Hence
\be
(a+b) \equiv \left( \frac{M_{a}X_{a}+M_{b}X_{b}}{M_{a}+M_{b}},P_{a}+P_{b},
M_{a}+M_{b} \right) \label{12}
\ee

We shall restrict our analysis to spatially homogeneous systems. 
The one-particle density does not depend in this case on the position
variable  
\be
\mu_{1}(1;t) = \mu_{1}(P_{1},M_{1};t) \label{13}
\ee
Therefore, the changes of $\mu_{1}(1;t)$ in the course of time are exclusively
due to collisions. Clearly, in one dimension only the nearest neighbours can 
collide.
The rate of collisions between adjacent particles $j$ and $(j+1)$ is 
proportional to the density of the nearest neighbours $\mu_{2}(j,j+1;t)$.
It can be written as
\be
C(j,j+1)\mu_{2}(j,j+1;t) \label{14}
\ee
where the collision factor
\be
C(j,j+1) \equiv \left( \frac{P_{j}}{M_{j}}-\frac{P_{j+1}}{M_{j+1}}\right)
\theta \left( \frac{P_{j}}{M_{j}}-\frac{P_{j+1}}{M_{j+1}} \right)
\delta (X_{j+1}-X_{j}-0+) \label{15}
\ee
chooses (with the help of the Dirac $\delta$-distribution) only the 
precollisional configurations. As the motion between collisions is free, the
rate (\ref{14}) is proportional to the relative velocity
of the  colliding pair (the $\theta$-factor in $C(j,j+1)$ assures the 
mutual approach of the particles). 

In order to evaluate the time derivative of the one-particle density 
(\ref{13}) let us consider first the events which lead to the creation of
the state $1$ by adhesive collisions. Their rate of occurrence is given
by 
\be
\int d1' \int d2'\; C(1',2')\delta[1-(1'+2')]\mu (1',2';t) \label{16}
\ee
Here the distribution 
\be
\delta[1-(1'+2')] = \delta \left[ X_{1}-
\frac{M_{1}'X_{1}'+M_{2}'X_{2}'}{M_{1}'+M_{2}'}\right] 
\delta [P_{1}-(P_{1}'+P_{2}')]
\delta [M_{1}-(M_{1}'+M_{2}')] \label{17}
\ee
picks out only those collisions which create the aggregate in the state
$1$. 

Consider now the events leading to the destruction of state $1$. When the 
precollisional state of one of the particles is $1$, the collision removes 
this state from the system. The rate of such annihilating events is given by
\be
\int d1' C(1',1)\mu_{2}(1',1;t) + \int d1' C(1,1')\mu_{2}(1,1';t)
\label{18}
\ee
Combining the gain term (\ref{16}) and the loss terms (\ref{18}), 
we arrive at the equation
\beq
\frac{\partial}{\partial t}\mu_{1}(1;t) &=&
\int d1' \int d2' C(1',2')\delta[1-(1'+2')]\mu_{2}(1',2';t)\non 
&&\quad -\int d1' C(1',1)\mu_{2}(1',1;t) - \int d1' C(1,1')\mu_{2}(1,1';t) 
 \label{19}
\eeq 

In a similar way one can derive the evolution equation for the two-particle
distribution of the nearest neighbours. It reads
\beq
&&\left[ \frac{\partial}{\partial t} + L_{12} + C(1,2) \right]\,
\mu_{2}(1,2;t)=\non
&&\quad \int d1' \int d2' C(1',2')\{ \delta[1-(1'+2')]\mu_{3}(1',2',2;t)+ 
\delta[2-(1'+2')]\mu_{3}(1,1',2';t)\}\non 
&&\quad\quad\quad
-\int d1' C(1',1)\mu_{3}(1',1,2;t) - \int d1'C(2,1')\mu_{3}(1,2,1';t) 
\label{20}
\eeq
where
\be
L_{12} = \frac{P_{1}}{M_{1}}\frac{\partial}{\partial X_{1}}+
         \frac{P_{2}}{M_{2}}\frac{\partial}{\partial X_{2}} \label{21}
\ee
is the generator of free streaming. On the left hand
side there appears the loss term $C(1,2)\mu_{2}(1,2;t)$ describing the 
destroying effect of a possible merging of the pair $(1,2)$. 
The right hand side takes into account the processes of creation of 
aggregates in the states $1$ or $2$ (the first two terms). Finally, the
last two terms represent the destruction of the two-particle state $(1,2)$
by collisions with the left nearest neighbour of particle $1$, and with 
the right nearest neighbour of particle $2$, respectively.
Equation (\ref{20}) relates the rate of change of $\mu_{2}(1,2;t)$ to the 
three-particle density $\mu_{3}(1,2,3;t)$. 

Denoting by $L_{1\ldots k}$ the k-particle generator of free streaming
\be
L_{1\ldots k} = \sum_{j=1}^{k}\frac{P_{j}}{M_{j}}\frac{\partial}
{\partial X_{j}}
\label{22}
\ee
we write the general $k$-th equation of the infinite hierarchy 
$(k=1,2,\ldots )$ in the form
\beq
&& \left[ \frac{\partial}{\partial t} + L_{1\ldots k} + 
\sum_{j=1}^{k-1}C(j,j+1) \right]\,\mu_{k}(1,2,\ldots ,k;t) =\non
&& \sum_{j=1}^{k} \int d1' \int d2' C(1',2') \delta[j-(1'+2')]
\mu_{k+1}(1,\ldots ,j-1,1',2',j+1,\ldots ,k;t) \non
&&\quad  -\int d1' C(1',1)\mu_{k+1}(1',1,2,\ldots ,k;t) - 
\int d1'C(k,1')\mu_{k+1}(1,2,\ldots ,k,1';t). 
\label{23}
\eeq
The fact that $\mu_{k}(1,\ldots ,j,j+1,\ldots ,k;t)$ 
is the density of a sequence
of $k$ nearest neighbours imposes a restrictive condition on its support.
Indeed, tracing backward in time the free trajectory of particle $j$  to the  
moment $t=0$, we find the position of the center of mass of the part of the 
initial system it has been formed from. So, if we denote by $\rho$ the 
constant mass density at $t=0$, the point
 \be
 X_{j}^{\rm right}(-t) = X_{j}-\frac{P_{j}}{M_{j}}t + \frac{M_{j}}{2\rho} 
 \equiv X_j-Y_j^-\label{26}
 \ee 
has the meaning of the right extremity of the initial mass region which 
produced the mass $M_{j}$ through adhesive collisions. This right extremity
must coincide with the left extremity
\be
X_{j+1}^{\rm left}(-t) = X_{j+1} - \frac{P_{j+1}}{M_{j+1}}t - 
\frac{M_{j+1}}{2\rho}\equiv X_{j+1}-Y_{j+1}^+ \label{27}
\ee
of the region which produced the mass $M_{j+1}$. Otherwise, the pair $(j,j+1)$
would not represent nearest neighbours.
In (\ref{26}) and (\ref{27}) we have introduced the notation
\be
Y^{\pm}_j=\frac{P_j}{M_j}t\pm\frac{M_j}{2\rho}.
\ee
We thus conclude that the 
density $\mu_{k}(1,\ldots ,k;t)$  contains necessarily the singular factor
\be
\prod_{j=1}^{k-1}\delta \left[ X_{j+1}-X_{j} -Y_{j+1}^++ Y_j^-\right]
\label{28}
\ee
This conclusion implies important consequences for collisional configurations 
where $X_{j+1}=X_{j}+0+$. The distribution (\ref{28}) imposes then the 
condition
\be
Y_{j+1}^+=Y_j^-
\label{29}
\ee
Hence, in the hierarchy equations (\ref{23}) one can replace the collision
terms $C(j,j+1)$ (see (\ref{15})) by
\be
\frac{M_{j}+M_{j+1}}{2\rho t}\delta (X_{j+1}-X_{j}-0+) \label{290}
\ee
With the use of (\ref{290}), one finds, after integrating over all momenta, 
the infinite hierarchy for the distribution of masses and distances between 
the particles which has already been derived in the study of the Burgers 
model of turbulence \cite{tat72}. 
Our independent reasoning based on the microscopic laws of the
aggregation process confirms the existence of a one-to-one correspondence 
between the two problems.

It is quite remarkable that the infinite set of equations (\ref{23}) 
can be rigorously reduced to a system of two coupled nonlinear equations 
under a simple assumption, which can be shown to be satisfied for 
a large class of initial conditions. This important fact, derived in the
next section, has not been noticed in \cite{tat72}. The authors restricted
their analysis therein to the first equation of the hierarchy,
supplemented with an integral relation obtained by integrating equation 
(\ref{20}) over both momenta and masses. In the search for 
self-similar solutions, they assumed a specific form of two-particle
correlations, which eventually led them to an erroneous conclusion
that the masses were exponentially distributed . The exponential
distribution has been eighteen years later conjectured again in \cite{pom90}, 
and shown to follow  from the hierarchy equations (\ref{23})
within the weak mean-field approximation \cite{jpias92}. It turned out that
a rigorous analysis invalidated these approximate results (see \cite{frach99}
and section \ref{sec9}).

\section{REDUCTION OF THE HIERARCHY}\label{sec4}

A fundamental role in further considerations will be played by the conditional
probability density 
\be
\mu (2|1;t) = \frac{\mu_{2}(1,2;t)}{\mu_{1}(1;t)} \label{24}
\ee
for finding  the right nearest neighbour of a particle supposed to occupy 
the state $1$, in the state $2$ at time $t$. The normalization 
\be
\int d2 \; \mu(2|1;t) = 1 \label{25}
\ee 
expresses the fact that in a homogeneous system the right nearest neighbour 
does exist in some state with certainty.

The singular factor
\be 
\delta \left[ X_{2}-X_{1} -Y_2^++Y_1^-\right]
\label{2900}
\ee
present in $\mu_{2}(1,2;t)$ (see (\ref{28})), and thus also in
$\mu(2|1;t)$, involves the combination $Y_1^-=tP_1/M_1-M_1/2\rho$ 
of mass and momentum of particle $1$.
This combination appears in formula (\ref{26}) for the position
of point $X_{j}^{\rm right}(-t)$, separating from the right hand side 
the masses which contributed to the formation of particle $j$ from the rest of 
the system. 

The reduction of the dynamical hierarchy (\ref{23}) presented in this section 
is possible if $\mu(2|1;t)$ depends on $P_{1}$ and $M_{1}$
exclusively via the effective variable $Y_1^-$. The following
structure of the conditional probability density will be thus assumed
\be
\mu(2|1;t) =\delta\left[ X_{2}-X_{1} -Y_2^++Y_1^-\right]
\bar{\mu }(M_{2},P_{2},Y_1^-;t) 
\label{00} 
\ee
In fact, it will be shown by construction in Section \ref{sec7} that for a
large class of initial conditions the above hypothesis is verified. 

The presence of factor (\ref{28}) in $\mu_{k}(1,\ldots ,k;t)$ clearly 
indicates the existence of correlations between the states of the nearest
neighbours. Let us suppose that only two-particle correlations exist, so
that 
\be
\mu_{k}(1,2,\ldots ,k;t) = \mu_{1}(1;t)\prod_{j=1}^{j=k-1}\mu (j+1|j;t)
\label{31}
\ee 
We shall prove now that the above factorization  is compatible with the
dynamics of the system, and that the factorized densities (\ref{31})
yield a solution to the hierarchy equations (\ref{23}), provided 
$\mu_{1}(1;t)$ and $\mu (2|1;t)$ satisfy the system of coupled equations
\beq
\frac{\partial}{\partial t}\mu_{1}(1;t) &=&   
\int d1' \int d2' C(1',2')\delta[1-(1'+2')]\mu_{1}(1';t)\mu (2'|1';t) \non
&&\quad  
-\int d1' C(1',1)\mu_{1}(1';t)\mu (1|1';t) - \int d1' C(1,1')\mu_{1}(1;t)
\mu (1'|1;t) 
\label{32} 
\eeq

\beq
\left[ \frac{\partial}{\partial t} + L_{12} + C(1,2) \right]\,&&\mu (2|1;t)
=  \int d1' \int d2' C(1',2') \delta[2-(1'+2')]\mu (1'|1;t)\mu (2'|1';t) 
\nonumber \\
&&\quad  +\left\{ \int d1' C(1,1')\mu (1'|1;t)  
- \int d1'C(2,1')\mu (1'|2;t)\right\} \mu (2|1;t) 
 \label{33} 
\eeq
Equation (\ref{32}) is just the first equation of the hierarchy (\ref{19})
in which $\mu_{2}(1,2;t)$ has been replaced by $\mu_{1}(1;t)\mu (2|1;t)$.

Let us consider the general equation (\ref{23}) for $k \geq 2$. 
Upon inserting the factorized form (\ref{31}) of the
densities $\mu_{k}$ and $\mu_{k+1}$ we get (with the use of the
first equation (\ref{32})) the following relation
\beq
&& \mu_{1}(1;t)\left[ \frac{\partial}{\partial t} + L_{1\ldots k} + 
\sum_{j=1}^{k-1}C(j,j+1)\right] \prod_{r=1}^{k-1}\mu (r+1|r;t) =\non
&& \int d1'\int d2' C(1',2')\delta[1-(1'+2')] \mu_{2}(1',2';t)[\mu (2|2';t)
- \mu (2|1;t)] \prod_{j=2}^{j=k-1}\mu (j+1|j;t) \non
&&\quad +\mu_{1}(1;t)\sum_{j=2}^{k}\int d1'\int d2'C(1',2')\delta[j-(1'+2')]
\prod_{r=1}^{j-2}\mu (r+1|r;t)  \non
&&\quad\quad\quad\quad 
\times  \mu (1'|j-1;t)\mu (2'|1';t)\mu (j+1|2';t)\prod_{s=j+1}^{k-1}
\mu (s+1|s;t) \non
&& \quad + \mu_{1}(1;t)\left[ \int d1' C(1,1')\mu (1'|1;t) - 
\int d1' C(k,1')\mu (1'|k;t)\right] \prod_{j=1}^{k-1}\mu (j+1|j;t).
\label{34} 
\eeq
The first term on the right hand side vanishes. Indeed, equation (\ref{29})
implies the equality 
\be
-\frac{P_{1}}{M_{1}}t + \frac{M_{1}}{2\rho}=  
-\frac{P_{1}'+P_{2}'}{M_{1}'+M_{2}'}t + \frac{M_{1}'+M_{2}'}{2\rho}=
-\frac{P_{2}'}{M_{2}'}t + \frac{M_{2}'}{2\rho}
\label{340}
\ee
So, taking into account the supposed structure (\ref{00}) of conditional 
densities, we can replace $\mu (2|2';t)$ in (\ref{34}) by $\mu (2|1;t)$.
Using exactly the same reasoning one can show that in the remaining 
gain terms in (\ref{34}), involving the creation of particles 
$j= 2,3,\ldots ,k$ 
through aggregation of $1'$ and $2'$, the density $\mu (j+1|2';t)$ 
equals $\mu (j+1|j;t)$. This permits to rewrite the hierarchy equation
(\ref{34}) in the following form 
\beq
&&\left[ \frac{\partial}{\partial t} + L_{1\ldots k} + 
\sum_{j=1}^{k-1}C(j,j+1)\right] \prod_{j=1}^{k-1}\mu (j+1|j;t) =\non
&& \sum_{j=2}^{k}\int d1'\int d2'C(1',2')\delta[j-(1'+2')]
\prod_{r=1}^{j-2}\mu (r+1|r;t)\mu (1'|j-1;t)\mu (2'|1';t)\prod_{s=j}^{k-1}
\mu (s+1|s;t) \non
&&\quad\quad + \left[ \int d1' C(1,1')\mu (1'|1;t) - 
\int d1' C(k,1')\mu (1'|k;t)\right] \prod_{j=1}^{k-1}\mu (j+1|j;t)
\label{35} 
\eeq
The last term in (\ref{35}) can be conveniently rewritten as
\be
\sum_{j=2}^{k}\left[ \int d1' C(j-1,1')\mu (1'|j-1;t) - 
\int d1' C(j,1')\mu (1'|j;t)\right] \prod_{r=1}^{k-1}\mu (r+1|r;t)
\label{36}
\ee
A straightforward calculation shows then that equation (\ref{34}) is satisfied
(for any $k\geq 1$) if the conditional probability density $\mu$ is a 
solution of equation (\ref{33}).

We have thus proved that solving the system of coupled equations
(\ref{32}) and (\ref{33}), one obtains the solution of the infinite
hierarchy (\ref{23}) in the factorized form (\ref{31}). This remarkable 
reduction of the dynamical hierarchy to a closed system of two equations
does not involve any approximation, provided $\mu (2|1;t)$ has the
structure (\ref{00}). The adhesive collisions do correlate the motion of
the nearest neighbours (the mean field approach is ruled out), but no three- 
or more-particle correlations are created in the course of time. The same 
kind of situation has been already discovered in the case of ballistic 
annihilation \cite{jpias95}.

\section{EVOLUTION OF REDUCED DENSITIES}\label{sec5}
% $\rho_{1} (P_{1},M_{1};t)$ AND $\rho
%(X_{21},P_{2},M_{2}| P_{1},M_{1};t)$ }

The factorized form (\ref{31}) of distributions $\mu_{k}$ implies the 
analogous factorization of the reduced densities
\be
\rho (1,\ldots ,k;t) = \rho_{1}(1;t)\prod_{j=1}^{j=k-1}\rho (j+1|j;t) 
\label{310}
\ee
where $\rho (2|1;t)$ is the conditional density
\be 
\rho (2|1;t) \equiv \rho (X_{21},P_{2},M_{2}|0,P_{1},M_{1};t)=
\frac{\rho _{2}(1,2;t)}{\rho_{1}(1;t)} \label{37}
\ee  
 From the closed system of equations (\ref{32}), (\ref{33}) for the nearest
neighbour distributions one can derive the corresponding set of dynamical 
equations for the densities $\rho_{1}=\mu_{1}$ and $\rho (2|1;t)$. First of 
all let us notice that the relation (\ref{9}) implies the equality
\be
\rho_{2}(X_{1},P_{1},M_{1},X_{1}+,P_{2},M_{2};t)
 = \mu_{2}(X_{1},P_{1},M_{1},X_{1}+,P_{2},M_{2};t) \label{38}
\ee
Hence, equation (\ref{32}) preserves its form
\beq
\frac{\partial}{\partial t}\rho_{1}(1;t) &=&   
\int d1' \int d2' C(1',2')\delta[1-(1'+2')]\rho_{1}(1';t)\rho (2'|1';t)\non 
&&\quad  -\int d1' C(1',1)\rho_{1}(1';t)\rho (1|1';t) - 
\int d1' C(1,1')\rho_{1}(1;t)\rho (1'|1;t) 
\label{39} 
\eeq
In order to derive the closed evolution equation for $\rho(2|1;t)$ one has
to use the formula 
\beq
\rho (2|1;t) &=& \mu (2|1;t) + \sum_{r=1}^{\infty}\int d1'\int d2' \ldots 
\int dr'\, \theta (X_{1}'-X_{1}) \theta (X_{2}-X_{r}') \non
&&\quad \times \prod_{j=1}^{r-1}\theta (X_{j+1}'-X_{j}') 
\,\mu (1'|1;t)\mu (2'|1';t)\cdots\mu (r'|r'-1;t)\mu(2|r';t) 
\label{40}
\eeq
(see (\ref{9}) and (\ref{31})).
Using then repeatedly equation (\ref{33}) one arrives at the equation
\beq
 \left[ \frac{\partial}{\partial t} + L_{12} + C(1,2)\right]\,&& \rho (2|1;t)
= \int d1' \int d2' C(1',2') \delta[2-(1'+2')]\rho (1'|1;t)\rho (2'|1';t) 
\nonumber \\
&&  + \int d1'C(1,1')\rho (1'|1;t) [ \rho (2|1;t)-\rho (2|1';t)]  
 \nonumber \\
&& -\int d1'\{ C(1',2)\rho (1'|1;t)\rho (2|1';t)+C(2,1')
\rho (1'|2;t)\rho (2|1;t) \} 
\label{41} 
\eeq 
Equations (\ref{39}) and (\ref{41}) form a closed system which suffices to
determine the reduced densities of any order owing to the relation (\ref{310}).

\section{SELF-SIMILAR SOLUTIONS}\label{sec6}

It seems interesting to check whether the aggregation dynamics is compatible
with self-similar solutions. In other words, whether the evolution of the
merging masses can be entirely reduced to rescaling of densities of a given
shape (this idea has been followed in the study  of the Burgers model of
turbulence \cite{tat72}). In order to investigate this question we insert into
equations (\ref{32}), (\ref{33}) the assumed scaling formulas
\beq
\mu_{1}(P_{1},M_{1};t) &=& t^{\gamma_1}\, 
\mu_{1}\(P_{1}t^{1-2\alpha},M_{1}t^{-\alpha};1\)
\label{42}\\
\mu (X_{21},P_{2},M_{2}|0,P_{1},M_{1};t) & = & t^{\gamma_2}\, 
\mu(X_{21}t^{-\alpha},P_{2}t^{1-2\alpha},M_{2}t^{-\alpha}|0,
P_{1}t^{1-2\alpha},M_{1}t^{-\alpha};1)
\label{43}
\eeq
where $X_{21}=X_{2}-X_{1}$ and where we have used the fact that
the three variables
\be
 X_{21}, \; \frac{P}{M}t,\; \frac{M}{2\rho} 
\ee
have the same dimension and  must thus scale in the same way. 

As the mass is conserved, the integral
\be
\int_{-\infty}^{+\infty} dP \int_{0}^{\infty} dM M\, t^{\gamma_1}\, 
\nu (Pt^{1-2\alpha},Mt^{\alpha}) \label{44}
\ee
representing the total  mass density does not depend on time. This implies the
relation
\be
\gamma_1 = 1-4\alpha\label{45}.
\ee
On the other hand, the normalization condition (\ref{25}) leads to
\be
\gamma_2 = 1-4\alpha\label{46}.
\ee

It can be checked by a straightforward calculation that the distributions 
$\mu_{1}(1;t)$ and $\mu (2|1;t)$ of the form (\ref{42}) and (\ref{43}), 
respectively, with the exponents $(\alpha, \gamma_1, \gamma_2 )$ satisfying
equations (\ref{45}) and (\ref{46}), lead to a consistent closed system of
equations for the self-similar distributions $\mu_1(P^\prime,M^\prime;1)$ 
and $\mu (2'|1';1)$
when put into the reduced hierarchy (\ref{32}), (\ref{33}). It follows that the
dynamics does not suffice to fix completely the values of the exponents.
They can thus depend on the nature of the initial condition of the system.

\section{A STATISTICAL MECHANICAL MODEL}\label{sec7}

In this section, we give a microscopic description of the dynamics of 
ballistic aggregation from the view-point of statistical mechanics by 
recalling the model introduced in \cite{pom90} and \cite{Mar-Pia}. 

One considers the aggregation process developing from a simple initial 
condition. At $t=0$, all the particles have the same
mass $m$ and are located on the sites $X_{j}=ja$
of an infinite regular lattice with lattice constant $a$ ($j=0,\pm 1 ,
\pm 2, ...$).  The mass density $\rho=m/a$. 
It is assumed that the initial momenta are uncorrelated, with a 
distribution corresponding to thermal equilibrium at inverse temperature 
$\beta$:
\be
\varphi_{m}(p)=\(\frac{\beta}{2\pi m}\)^{1/2}\exp\(-\frac{\beta p^{2}}{2m}\)
\label{6.1}
\ee
As it has already been explained in Section \ref{sec3}, the dynamics of
the sticky gas has the remarkable property that any mass aggregate  
is found on the trajectory of the center of mass of the initial cluster it
has been formed from. In particular, the state of an aggregate
$M_{j}=n_{j}m$ at time $t > 0$ determines uniquely the set of 
$n_{j}$ consecutive initial masses which constitute it: they are located
at $t=0$ within the interval $\left[X_j-Y_j^++a/2,X_j-Y_j^--a/2\right]$ 
(see also Eqs.(\ref{26}),(\ref{27})).
This property permits to write explicitly the kinematical constraints that
select the subset of initial phase space configurations leading to the 
occurrence of a specific sequence of aggregate states 
$j = (X_{j},P_{j},M_{j}),\;j=1,\ldots,k,$ within $[L_{1},L_{2}]$ at time 
$t>0$. The distributions $\mu_{k}(1,2,\ldots,k;t\,|L_{1},L_{2})$ are then
obtained by averaging these constraints over the initial state.

The origin of the constraints is two-fold. One class of them ensures that
the $n_{j}$ initial particles located in the interval 
$[X_j-Y_j^++a/2,X_j-Y_j^--a/2]$ do merge and form the $j^{th}$ aggregate
before time $t$. A necessary and sufficient condition here 
is that the trajectories of the centers of mass of all the pairs of 
subclusters of consecutive initial particles, which form a partition of
the initial $n_{j}$-particle cluster, 
%located in 
%$[X_j-Y_j^++a/2,X_j-Y_j^++a/2+ai]$ and $[X_j-Y_j^++ai+3a/2,X_j-Y_j^--a/2]$, 
%$i=1,\ldots, n_j$,
cross before time $t$. After averaging, this leads to functions denoted by
$I_{m}(M_j,P_j;t)$ which are the probability
densities for the formation of masses $M_j=n_jm$, with momenta $P_j$ 
from the corresponding sets of $n_j$ initial neighbouring masses.

The second class of constraints guarantees that no particles other than 
the $k$ specified aggregates are found within $[L_{1},L_{2}]$ at time $t$. 
The particles initially located in $(-\infty,ja]$ will be found to the left of
$L_{1}$ if the positions of all the centers of mass of clusters of 
consecutive initial masses located in $[(j-i)a,ja]$, $i=1,2,\ldots$ stay 
smaller than $L_{1}$ up to time $t$.
After averaging, this will be expressed by  a function
$J_{m}(Y;t)$, where $J_{m}(-L_{1}+(j+1/2)a;t)$  (resp.  
$J_{m}(L_{2}-(j+1/2)a;t)$ is the probability for finding the particles 
initially located in  $(-\infty,ja]$ (resp. $[(j+1)a,\infty)$)
in $(-\infty,L_{1}]$ (resp. in $[L_{2},\infty)$). 
In particular, in terms of function $J_{m}$ defined in this way the 
probability $\mu_{0,m}(t|L_{1},L_{2})$ of finding the interval 
$[L_{1},L_{2}]$ void of particles is given by 
\be
\mu_{0,m}(t|L_{1},L_{2})=\sum_{j}J_m\(-L_{1} +(j+1/2)a;t\)
J_m\(L_{2} -(j+1/2)a;t\)
\label{6.2}
\ee
The probability density $\mu_{1,m}$ (\ref{6}) reads \footnote{We add the 
index $m$ to keep in mind the discrete nature of the initial state.}
\beq
\mu_{1,m}(1;t)=\sum_{n_{1}=1}^{\infty}\delta(M_{1} && -n_{1}m)\sum_{j}
\delta\(X_1-Y^{+}_{1}-(j+1/2)a\)\non
&&\times I_{m}(M_1,P_{1};t)J_{m}\(-Y^{+}_{1};t\)J_{m}\(Y^{-}_{1};t\)
\label{6.3}
\eeq
where the summation involving $\delta$-functions is a manifestation of 
the discreteness of the initial masses and the lattice positions. 
The function $I_{m}$ in (\ref{6.3}) ensures that the particles initially 
located in $[X_1-Y_1^++a/2,X_1-Y_1^--a/2]$ have met to form the aggregate,
while the functions $J_{m}$ ensure that all the particles initially
located on the left 
of $X_1-Y_1^+$ (resp. on the right of $X_1-Y_1^+$) are at time $t$ on the 
left (resp. on the right) of $X_1$.

The probability density $\mu_{2,m}$ (\ref{6}) is given by
\beq
\mu_{2,m}(1,2;t) &=& \theta(X_{2}-X_{1})
\sum_{n_{1}=1}^{\infty}\delta(M_{1}-n_{1}m)
\sum_{n_{2}=1}^{\infty}\delta(M_{2}-n_{2}m)\nonumber\\
&&\quad \times\sum_{j}\delta\(X_{1}-Y^{+}_{1}-(j+1/2)a\)
  \delta\(X_{2}-X_{1}+Y^{-}_{1}-Y^{+}_{2}\)\non
&&\quad\times 
I_{m}(M_1,P_{1};t)I_{m}(M_2,P_{2};t)
J_{m}\(-Y^{+}_{1};t\)J_{m}\(Y^{-}_{2};t\).
\label{6.4}
\eeq
In this expression the $n_i$ and $j$ summations involving $\delta$-functions 
reflect again the discreteness of the initial conditions.
The additional $\delta-$function obtained here by construction reflects 
the fact that the aggregates are nearest neighbours and imposes precisely
the condition (\ref{00}) under which the hierarchy could be reduced.
Once again, functions $I_{m}$ stand for the formation of aggregates $1$
and $2$, while functions $J_{m}$ ensure that
particles initially located on the left of $X_1-Y_1^+$ (resp. on the right
of $X_2-Y_2^-$) stay at time $t$ on the left of $X_1$ (resp. on the right
of $X_2$).

The higher order distributions $\mu_{k,m}$ can be constructed along the same 
lines. They verify the factorization 
(\ref{31})
\be
\mu_{k,m}(1,\ldots,k;t)=\mu_{1,m}(1;t)\prod_{j=1}^{k-1}\mu_{m}(j+1|j;t)
\label{6.5}
\ee
with the conditional probability $\mu_{m}(2|1;t)$, derived 
from (\ref{6.3}) and (\ref{6.4}), of the form
\beq
&&\mu_{m}(2|1;t)=\frac{\mu_{2,m}(1,2;t)}{\mu_{1,m}(1;t)}\non
&&\quad =\theta(X_{2}-X_{1})\sum_{n_{2}=1}^{\infty}\delta(M_{2}-n_{2}m)
\delta\(X_{2}-X_{1}+Y^{-}_{1}-Y^{+}_{2}\)I_{m}(M_2,P_{2};t)
\frac{J_{m}\(Y^{-}_{2};t\)}{J_{m}\(Y^{-}_{1};t\)}
\label{6.6}
\eeq
Note that this conditional probability has the structure (\ref{00}).

\begin{figure}
%\narrowtext
\epsfxsize=11truecm
\hspace{2.75truecm}
\epsfbox{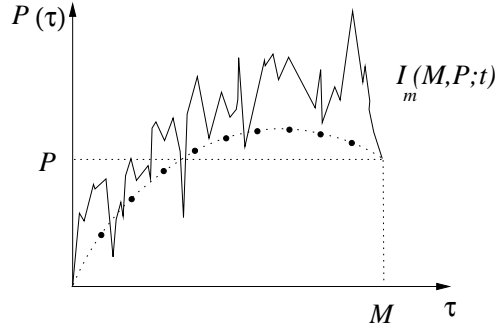}
\caption{Brownian interpretation of the function $I_m(M,P;t)$. The Brownian 
motion starts at $P(0)=0$, ends at $P(M)=P$ while overpassing the points
$P(rm)\geq rm(2\rho Q^+-rm)/2\rho t$, ($r=1,\ldots,n-1$) 
with $Q^+=Pt/M+M/2\rho$.}
\label{fig1}
\end{figure}

To have the model in an explicit form, it remains to give the formulae 
that express functions $I_{m}$ and
$J_{m}$ in terms of the constraints: the details of the calculation can be 
found in \cite{Mar-Pia}. 
The important point is that, due to the uncorrelated Gaussian initial 
velocity distribution, the whole aggregation dynamics can be mapped
on the following equivalent Brownian motion problem:
 
Let $P(\tau)$ be a Brownian path starting from the origin, $P(0)=0$. 
Then, for $M=nm$, 
\be
I_{m}(M,P;t)=
E_{(0,0)}\{ P(rm)\geq rm[(M-rm)/2\rho t+P/M],r=1,\ldots,n-1|P(nm)=P\} 
\label{6.7}
\ee
is the conditional measure of such paths constrained to 
be above the parabolic barrier $rm[(M-rm)/2\rho t+P/M]$
at discrete "times" $\tau_{r}=rm,\;r=1,\ldots,n-1$ and to 
end in $P$ at "time" $nm$, see Fig.\ref{fig1}. 

\begin{figure}
%%\narrowtext
\epsfxsize=11truecm
\hspace{2.75truecm}
\epsfbox{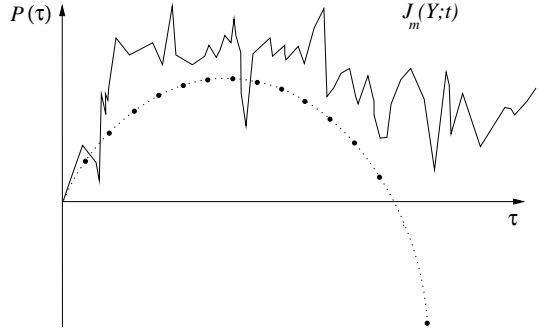}
\caption{Brownian interpretation of the function $J_m(Y;t)$. 
The Brownian motion starts at $P(0)=0$ and overpasses the points 
$P(rm)\geq rm(2\rho Y-rm)/2\rho t$, ($r=1,2,\ldots$).}
\label{fig2}
\end{figure}

Likewise
\be
J_{m}(Y;t)=E_{(0,0)}\{ P(rm)\geq {rm(2\rho Y-rm)}/{2\rho t},r=1,2,\ldots\} 
\label{6.8}
\ee
is the measure of the paths that remain above the barrier
${rm(Y-rm)}/{2\rho t}$ for all discrete times
$\tau_{r}=rm,\;r=1,\ldots$; see Fig.\ref{fig2}. We recall that the dynamics
is deterministic, and randomness enters only through the initial velocity
distribution: $P(\tau)$ is a process in 
momentum space, as a function of the mass of aggregates.

A simple and immediate consequence are the scaling properties of functions
$I_{m}$ and $J_{m}$ with respect to time. Owing to the fact that the
scaled Brownian motion
$(\rho t)^{\alpha/2}P(\tau (\rho t)^{-\alpha})$ 
is equivalent in probability to $P(\tau)$, 
the functions $I_{m}$ and $J_{m}$ obey the scaling relations with 
$\alpha=2/3$ 
\beq
&&I_{m}(M,P;t)=(\rho t)^{-1/3}I_{m^{\prime}}(M^\prime,P^{\prime};1)\equiv
(\rho t)^{-1/3}I_{m^{\prime}}(M^\prime,P^{\prime})\nonumber\\ 
&&J_{m}(Y;t)=J_{m^{\prime}}(Y^{\prime};1)\equiv J_{m^{\prime}}(Y^{\prime})
\label{6.9}
\eeq
with
\be
m^{\prime}=m(\rho t)^{-2/3},\;\;P^{\prime}=P(\rho t)^{-1/3},\;\;
 M^{\prime}=nm^{\prime}=M(\rho t)^{-2/3}
,\;\;Y^{\prime}=(\rho t)^{-2/3}\rho Y
\label{6.10}
\ee
The construction (\ref{6.3})-(\ref{6.8}) yields in principle a 
solution to the dynamical hierarchy 
that has the factorization property: $\mu_{1,m}(1;t)$ and 
$\mu_{m}(2|1;t)$ have to verify the coupled equations
(\ref{32})-(\ref{33}), adapted to the case of discrete masses. 
We shall not proceed to this verification now, but rather simplify first
the discussion by taking the continuum limit of our statistical mechanical
model.

The continuum limit amounts to let $m\to 0$ and $a\to 0$ while 
keeping the initial mass density $\rho=m/a$ and time fixed. Because of 
(\ref{6.9}) one can alternatively fix the scaled momentum $P^{\prime}$
and the scaled mass $M^{\prime}$ and look for the large time asymptotics 
$t\to\infty$. In the first view, one looks for the formation of masses of 
order $1$ at a given time arising from an initial infinitesimal
dust. In the second view one looks for the formation of 
masses of size $\sim mt^{2/3}$, $t\to\infty$, while keeping the discrete 
initial condition. These two views are equivalent. The interesting point
is that in the continuum limit the Brownian expressions (\ref{6.7}) and 
(\ref{6.8}) can be computed in a closed analytical form, thus
providing an explicit complete statistical description of the aggregation
process. Moreover this description exactly  coincides with that of the 
statistics of shocks in the inviscid limit of the Burgers equation with 
white noise initial data.

To conclude one can check that the factorization property (\ref{6.5})
persists for a wider class of initial conditions, for instance 
allowing any uncorrelated initial velocity distribution, and an arbitrary
(non random) choice of initial positions and masses.

\section{THE CONTINUUM LIMIT}\label{sec8}

In this section, we determine the continuum limits 
($m\to 0$ with $\rho=m/a$ fixed) of the one- and two-point 
probability densities
\beq
&&\lim_{m\to 0} \mu_{1,m}(1;t)=\mu_1(1;t)\non
&&\lim_{m\to 0}\mu_{2,m}(1,2;t)=\mu_2(1,2;t).
\label{7.0}
\eeq
Owing to the scaling relations (\ref{6.9}) it is sufficient to calculate 
functions $\mu_1(1;t)$ and $\mu_2(1,2;t)$ for $t=1$ 
and\footnote{In \cite{Mar-Pia} the density was set equal to $1/2$.} 
$\rho=1$. 
From now on we drop the time parameter from the notation
setting simply $\mu_{1,m}(1;t)=\mu_{1,m}(1)$, 
$\mu_{2,m}(1,2;t)=\mu_{2,m}(1,2)$, and so on. Later in this section
we will comment on the way to recover the time variable in the continuum
limit of these densities.

Let us first introduce the transition probability density kernel
\be
K_{m,\nu}(M_1,P_1,M_2,P_2)=\quad E_{(M_1,P_1)}
\{ P(rm)\geq f_\nu(rm),r=n_1+1,\ldots,n_2-1|P(M_2)=P_2\} 
\label{7.1}
\ee
for the Brownian motion $P(\tau)$ to start at $P(M_1)=P_1$, and end at 
$P(M_2)=P_2$ with $M_1=n_1m$ and $M_2=n_2m$, while overpassing the points 
$f_\nu(rm)$, $r=n_1+1,\ldots,n_2-1$, where $f_\nu(\tau)$ is the parabola
\be
f_\nu(\tau)=\nu \tau-\frac{\tau^{2}}{2}.
\label{7.2}
\ee
According to definitions (\ref{6.7}) and (\ref{6.8}), 
the functions $I_m$ and $J_m$ appearing in the one- and two-point densities
can be expressed in terms of the transition kernel (\ref{7.1}) as 
\beq
I_m(M,P)= K_{m,Y^+}(0,0,M,P)
%&=& \int_{f_{Z^+}({\overline M},t)}^{\infty} 
%d{\overline P}\,K_{m,Z^+}(0,0,\overline{M},\overline{P};t) 
%K_{m,Z^-}(0,0,M-\overline{M},\overline{P}-P;t)
\label{7.3}
\eeq
 and 
\be
J_m(Y)=\lim_{M\to\infty}\int_{f_Y(M)}^\infty dP\, K_{m,Y}(0,0,M,P).
\label{7.4}
\ee

In (\ref{7.1}), Brownian paths are allowed to make excursions through
holes of width $m$ separating the discrete points $rm$.
In the continuum limit $m\to 0$, the weight of such excursions becomes 
vanishingly small and the paths become constrained to overpass the 
continuous barrier $f_\nu(\tau)$, $M_{1}\leq \tau\leq M_{2}$.  
Thus for $P_{1}>f_{\nu}(M_{1})$ and $P_{2}>f_{\nu}(M_{2})$
\be
K_{m,\nu}(M_1,P_1,M_2,P_2)= K_{\nu}(M_1,P_1,M_2,P_2) 
+R_{m,\nu}(M_1,P_1,M_2,P_2) 
\label{7.5}
\ee
where $\lim_{m\to 0}R_{m,\nu}(M_1,P_1,M_2,P_2)=0$, and
\be
K_\nu (M_1,P_1,M_2,P_2)
=E_{(M_1,P_1)}
\{P(\tau)> f_\nu (\tau),M_1\leq \tau\leq M_2|P(M_2)=P_2\} 
\label{7.8}
\ee
is the transition kernel for a Brownian motion with a continuous 
parabolic barrier.
It satisfies the diffusion equation 
\be
\frac{\partial}{\partial M_{2}}K_\nu (M_1,P_1,M_2,P_2)
=\frac{1}{2\beta}\frac{\partial^{2}}
{\partial P_{2}}K_\nu (M_1,P_1,M_2,P_2)
\label{7.7}
\ee
with $K_\nu (M,P_1,M,P_2)= \delta(P_{1}-P_{2})$, and the Dirichlet conditions 
on the barrier:
$K_\nu (M_1,P_1,M_2,P_2)=0$ when $P_{1}=f_{\nu}(M_{1})$ or  
$P_{2}=f_{\nu}(M_{2})$. 

Our object now is to express the distributions of aggregates in the
continuum limit in terms of the kernel (\ref{7.8}), which can be
explicitly computed (Section II of \cite{FraMar99}).
However, in the continuum limit of $I_m$ and $J_m$ the starting point $(0,0)$ 
of the Brownian motion lies on the parabola (\ref{7.2}) where
$K_{\nu}(0,0,M,P)=0$. Thus the determination of the densities $\mu_{1}(1)$ 
and $\mu_{2}(1,2)$ requires the evaluation of the leading term in the 
asymptotic expansion of $K_{m,\nu}(0,0,M,P)$ as $m\to 0$. This is not an
elementary task since it involves the
control of the cumulated effect of Brownian excursions in small
intervals $((r-1)m,rm),\;r=1,2,\ldots$ in the neighbourhood of the origin.
The result is given in the following proposition

\vspace{5mm}
\noindent
{\bf Proposition}

\be
\lim_{m\to 0}{K_{m,\nu}(0,0,M,P)\over \sqrt{m}}
={1\over\sqrt{2\beta}}\left.{\partial\over \partial P_0}
K_\nu(M_0,P_0,M,P)\right|_{(M_0,P_0)=(0,0)},\;M>0,\; P\geq f_\nu(M)
\label{7.8p}
\ee
\vspace{5mm}

With this result the continuum limits $m\to 0$ of functions $I_m$ and $J_m$, 
as expressed in (\ref{7.3}) and (\ref{7.4}), with fixed initial mass
density $\rho=m/a$, read
\be
\lim_{m\to 0}{I_m(M,P)\over m}={1\over 2\beta}
\left.{\partial^2\over
\partial P_1\partial P_2}K_{Y^+}(0,P_1,M,P_2)\right|_{P_1=0,P_2=P}
\equiv I(M,P)
\label{7.9}
\ee
and
\be
\lim_{m\to 0}{J_m(Y)\over\sqrt{m}}=
\lim_{M\to\infty}{1\over \sqrt{2\beta}}\int_{f_Y(M)}^\infty dP\,
\left.{\partial\over\partial P_1}K_{Y}(0,P_1,M,P)\right|_{P_1=0}
\equiv J(Y).
\label{7.10}
\ee  
In obtaining (\ref{7.9}) we have taken into account that in (\ref{7.3}),
in addition to $(0,0)$, the paths have a second contact point with
the parabola at $(M,P)$.

Then, using equations (\ref{6.3}),(\ref{6.4}), the continuum limits of the
one- and two-point probability densities can be readily found. One multiplies
and divides (\ref{6.3}) by $m^{2}$ and takes (\ref{7.9}) and (\ref{7.10})  
into account. Then,  $m\to 0$ playing the role of an infinitesimal, 
the discrete sums go to the corresponding integrals, which
can be evaluated owing to the $\delta$-functions, yielding eventually the 
formula
\be
\mu_1(1)= I(M_1,P_1)J\(-Y^{+}_{1}\)J\(Y^{-}_{1}\)
\label{7.11}
\ee
and, in the same way
\be
\mu_2(1,2)=  \theta(X_{2}-X_{1})
\delta\(X_{2}-X_{1}+Y^{-}_{1}-Y^{+}_{2}\)I(P_{1},M_{1})I(P_{2},M_{2})
J\(-Y^{+}_{1}\)J\(Y^{-}_{2}\).
\label{7.12}
\ee

The time and the initial density dependence can always 
be reintroduced in the continuum limit. Indeed, through 
Eqs.(\ref{6.9}) and (\ref{7.9}),(\ref{7.10}) we get the scaling 
properties of functions $I$ and $J$
\be
J(Y;t)=(\rho t)^{-1/3}J(Y^\prime),\quad I(M,P;t)=(\rho t)^{-1}
I(M^\prime,P^\prime ) \label{7.100}
\ee
with the scaling functions $J(Y)$ and $I(M,P)$ defined above. This 
implies the scaling behavior of the densities in the continuum limit of 
the form
\be
\lim_{m\to 0}\mu_{k,m}(1,\ldots,k;t)=
\mu_k(1,\ldots,k;t)=\rho^k (\rho t)^{-5k/3}\mu_k(1^\prime,\ldots,k^\prime)
\label{7.12b}
\ee
with 
\be
j^\prime\equiv (X_j^\prime,P_j^\prime,M_j^\prime)
=\({\rho X_j\over (\rho t)^{2/3}},{P_j\over (\rho t)^{1/3}},
{M_j\over (\rho t)^{2/3}}\).
\label{7.12c}
\ee 
The scaling functions $\mu_1(1)$ and $\mu_2(1,2)$ are given by
(\ref{7.11}) and (\ref{7.12}) and, for higher orders, the scaling functions
are given by
\be
\mu_k(1,\ldots,k)=\mu_1(1)\prod_{j=1}^{k-1}{\mu_2(j,j+1)\over \mu_1(j)}.
\label{7.13.0}
\ee

From Eq.(\ref{6.2}), we find that the probability density to find an 
interval $[0,x)$ void of aggregates scales in the continuum limit as
\be
\mu_0(x;t)=\mu_0(x^\prime )
\ee
with $x^\prime =\rho x (\rho t)^{-2/3}$ and 
\be
\mu_{0}(x)=\int dy J(y)J(x-y).
\label{7.13}
\ee
Since the functions $I(M,P)$ and $J(Y)$ have been explicitly computed 
in \cite{FraMar99}, the results (\ref{7.11},\ref{7.12}) and (\ref{7.13.0}) 
give a complete
solution of the ballistic aggregation model in a closed analytical form. 
We postpone the discussion of this solution to the next section
and devote the rest of the present section to the proof of (\ref{7.8p}).

\vspace{5mm}
\noindent
{\bf Proof of the proposition}

The strategy to prove Eq.(\ref{7.8p}) 
is to bound the transition kernel (\ref{7.1}) while linearizing the
constraints imposed on the first $k=M_0/m$ points, $0<M_{0}<M$. 

\begin{figure}
%\narrowtext 
\epsfxsize=11truecm
\hspace{2.75truecm}
\epsfbox{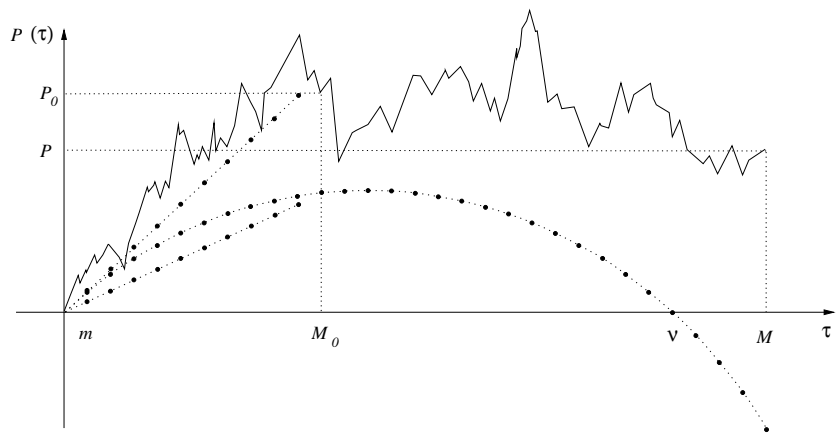} 
\caption{Illustration of the ``linearized'' bounds used to in 
(\protect\ref{7.17}).}
\label{fig3} 
\end{figure}

We have 
\be 
\(\nu-\frac{M_{0}}{2}\)\tau
\leq f_{\nu}(\tau)\leq \nu \tau,\;\,\,0\leq \tau\leq M_{0}
\label{7.14}
\ee
Let us introduce the kernel
\be
K^{l}_{m,\nu}(0,0,M_0,P_0)=
E_{(0,0)}\{ P(rm)\geq \nu rm,r=1,\ldots,k-1
|P(M_0)=P_0\}
\label{7.15}
\ee
for paths starting at $(0,0)$, ending at $(M_{0},P_{0})$, while overcoming
the linearly distributed sequence of discrete points $\nu rm$, $r=1,
\ldots,k-1$. Clearly (see Fig. \ref{fig3})
\be
 K^{l}_{m,\nu}(0,0,M_0,P_0)\leq
K_{m,\nu}(0,0,M_0,P_0)\leq K^{l}_{m,\nu-M_0/2}(0,0,M_0,P_0).
\label{7.16}
\ee

Using the Markov property of the Brownian motion, one infers from (\ref{7.16})
the bounds for the kernel $K_{m,\nu}(0,0,M,P)$ for $M>M_{0}>0$ 
\be
K_{m,\nu}^{(<)}(0,0,M,P)\leq K_{m,\nu}(0,0,M,P)\leq 
K_{m,\nu}^{(>)}(0,0,M,P)
\label{7.17}
\ee
with 
\be
K_{m,\nu}^{(<)}(0,0,M,P)=\int dP_0\,K^{l}_{m,\nu}(0,0,M_0,P_0)
K_{m,\nu}(M_0,P_0,M,P)
\label{7.18}
\ee
and
\be
K_{m,\nu}^{(>)}(0,0,M,P)=\int dP_0\,
K^{l}_{m,\nu-M_0/2}(0,0,M_0,P_0)
K_{m,\nu}(M_0,P_0,M,P).
\label{7.19}
\ee
We shall calculate these bounds by first letting $m\to 0$  
and then  $M_{0} \to 0$. They will be shown to coincide in this limit. 

It is convenient to relate the transition kernel $K^{l}_{m,\nu}$ 
(\ref{7.15}) for 
the Brownian  process with discrete linear barrier $ P(rm)\geq \nu rm$,
$r=1,\ldots,k-1$ to that of a dimensionless Brownian process $q(\tau)$ 
with covariance equal to $1$, the paths constrained to 
be positive at integer times, and  with the corresponding 
transition kernel
\be
G_{k}(q)=E_{(0,0)}\left\{q(n)\geq 0,n=1,\ldots k-1|q(k)=q\right\}.
\label{7.20}
\ee
For this later process, we define the average 
$\left\langle f(q)\right\rangle_{k}=\int dq\,G_{k}(q)f(q) $.
The moments $\left\langle q^{j}\right\rangle_{k},j=0,1,2,\ldots$ will be
for us of particular interest.

The relation between the two processes is given by removing a linear drift
\be
q(\tau)=\sqrt{\frac{\beta}{2m}}(P(\tau)-\nu \tau)
\ee
which leads to the relation 
\be
K^{l}_{m,\nu}(0,0,M_{0},P_{0})=\exp\(\frac{\beta\nu^{2}M_{0}}{2}\)
\sqrt{\frac{\beta}{2m}}G_{k}\(\sqrt{\frac{\beta}{2m}}(P_{0}-\nu M_{0})\),\;\,
k=\frac{M_{0}}{m}.
\label{7.21}
\ee
When this is inserted into (\ref{7.18}), one obtains
\beq
K_{m,\nu}^{(<)}(0,0,M,P) &=& \exp\(-\frac{\beta\nu^{2}M_{0}}{2}\)\non
&&\quad\times \int dq\,G_{k}(q){\rm e}^{-\nu(2m\beta)^{1/2}q}
K_{m,\nu}\(M_0,\nu M_0+\sqrt{{2m\over\beta}}q,M,P\).
\label{7.22}
\eeq
Then, we take the following steps. First, for $M_{0}>0$ we can replace
in (\ref{7.22}) the kernel $K_{m,\nu}$ for the discrete parabolic barrier
by its continuous limit (\ref{7.5}). Next, we expand the integrand up to
second order in the variable $\sqrt{m}q$. This gives  
\beq 
&& K_{m,\nu}^{(<)}(0,0,M,P)= {\rm e}^{-\beta \nu^2 M_0/2}
\biggr[\left\langle  1\right\rangle_{k} 
K_{\nu}(M_0,\nu M_0,M,P)\non
&&\left. \quad +\sqrt{m}\left\langle  q\right\rangle_{k}
\left(-\nu\sqrt{2\beta}K_{\nu}(M_0,\nu M_0,M,P)
+\sqrt{{2\over\beta}}{\partial\over\partial P_0}
K_{\nu}(M_0,P_0,M,P)|_{P_0=\nu M_0}\right)\right]\non
&&\quad\quad\quad\quad\quad\quad\quad\quad\quad
+\left\langle q^{2}\right\rangle_{k}{\cal O}(m) 
+\left\langle  1\right\rangle_{k} o(m).
\label{7.23}
\eeq 
Recalling that $k=M_0/m$, we have to evaluate the first moments 
$\left\langle q^{j}\right\rangle_{k}$
of the distribution $G_{k}(q)$ for large $k$. This is done in appendix 
\ref{app1}
with the help of the Sparre-Andersen theorem together 
with a Tauberian theorem. One gets
\be
\left\langle  1\right\rangle_{k}\sim\sqrt{\frac{M_{0}}{\pi m}},\;\;
\left\langle q\right\rangle_{k}\sim\frac{1}{2},\;\; \left\langle q^{2}\right
\rangle_{k}\sim\sqrt{\frac{m}{\pi M_{0}}}
,\;\;\;m\to 0.
\label{7.24}
\ee
Hence
\beq
&&\lim_{m\to 0}\frac{ K_{m,\nu}^{(<)}(0,0,M,P)}{\sqrt{m}}
={\rm e}^{-\beta \nu^2 M_0/2}\left[\frac{K_{\nu}(M_0,\nu M_0,M,P)}{\sqrt{\pi M_{0}}}
\right.\non &&\left.+\frac{1}{2}\left(-\nu\sqrt{2\beta}K_{\nu}(M_0,\nu M_0,M,P)
+\sqrt{{2\over\beta}}{\partial\over\partial P_0}
K_{\nu}(M_0,P_0,M,P)|_{P_0=\nu M_0}\right)\right]+{\cal O}(\sqrt{M_{0}})
\label{7.25}
\eeq
Finally, since $K_{\nu}(M_0,P_{0},M,P)$ is differentiable with respect to 
$P_{0}$
and vanishes at $P_{0}=0$, one has $K_{\nu}(M_0,\nu M_0,M,P)={\cal O}
(M_{0})$, $M_{0}\to 0$,  so that 
\be
\lim_{M_0\to 0}\lim_{m\to 0}{K_{m,\nu}^{(<)}(0,0,M,P)\over \sqrt{m}}=
{1\over \sqrt{2\beta}}\left.{\partial\over\partial P_0}
 K_{\nu}(M_0,P_0,M,P)\right|_{(M_0,P_0)=(0,0)}.
\ee
The upper bound $K_{m,\nu}^{(>)}(0,0,M,P)$ is treated in the same way
and attains the same limit, thus 
leading to the result Eq.(\ref{7.8p}) of the proposition.

The analysis performed in \cite{Mar-Pia} was based on the lower bound
\be
{K_{m,\nu}(0,0,M,P)\over \sqrt{m}}>
{1\over \sqrt{2\pi\beta}}\left.{\partial\over\partial P_0}
 K_{\nu}(M_0,P_0,M,P)\right|_{(M_0,P_0)=(0,0)}
\ee
obtained by retaining the effects of Brownian excursions only in the
first interval $(0,m)$ at the origin.
The result obtained therein differs from the exact limit (\ref{7.8p}) by a
factor  $1/\sqrt{\pi}$.
In fact, the limit (\ref{7.8p}) involves contributions of infinitely
many intervals, which can be summed up by applying the Sparre-Andersen 
theorem.

\section{Explicit solution}\label{sec9}

The function $I(M,P)$ (\ref{7.9}),
explicitly computed in \cite{FraMar99}, has the form
\be
I(M,P)=2b^3
\exp\left(-b^3\left[{P^2\over M}+{M^3\over 12}\right]\right)
{\cal I}(M),
\label{iii}
\ee
where we set $b=(\beta/2)^{1/3}$ and
\be
{\cal I}(M)=\sum_{k\geq 1} {\rm e}^{-b\omega_k M}.
\label{cali}
\ee
The function $J(Y)$ (\ref{7.10}) reads
\be
J(Y)=\sqrt{b}{\rm e}^{-b^3Y^3/3}{\cal J}(Y)
\label{jjj}
\ee
with
\be
{\cal J}(Y)={1\over 2i\pi }\int_{-i\infty}^{+i\infty}dw
{{\rm e}^{bYw}\over {\rm Ai}(w)}.
\label{calj}
\ee 
In (\ref{cali}) and (\ref{calj}), $\Ai(w)$ is
the Airy function \cite{Abra},  solving the differential equation
\be
f''(w)-wf(w)=0,
\ee
$\Ai(w)$ is analytic in the complex $w$ plane, and 
has an infinite countable set of zeros $-\omega_k$ on the 
negative real axis, $0<\omega_{1}<\omega_{2}<\cdots$. 
The asymptotic behavior of these functions has been derived in \cite{FraMar99}
\be
{\cal I}(\mu)\sim {1\over \sqrt{4\pi b^3\mu^3}},\;\;\mu\to 0;\quad
{\cal I}(\mu)\sim \exp(-\omega_1 b\mu),\;\;\mu\to \infty
\label{ias}
\ee
and
\be
{\cal J}(u)\sim {{\rm e}^{-bu\omega_1}\over {\rm Ai}'(-\omega_1)},\;\; 
u\to \infty;\quad
{\cal J}(u)\sim -2bu\exp\({b^3u^3\over 3}\),\;\; u\to -\infty.
\label{jas}
\ee

In fact, these results, which have been announced in \cite{frach99}, 
were derived 
in details in the related framework of the Burgers turbulence
\cite{FraMar99} . 
The solutions of the one-dimensional Burgers equation with a white-noise
initial condition, develop, in the limit of vanishing viscosity, 
a train of shock waves.
In Burgers theory, a shock located at $X$ is 
characterized by two parameters $\mu$ and $\eta$, and can be
identified with a particle of mass $M=\mu$ at point $X$  with
momentum \footnote{In Burgers language, one speaks of the
shock strength $\mu/t$ and shock wavelength $\nu=\mu/2-t\eta/\mu$} 
$P=-\eta$.
Then it turns out that dynamics of shocks is
completely equivalent to ballistic aggregation subject to mass and 
momentum conservation \cite{tat72},\cite{burgers}.
Moreover, the white noise covariance is $D/2=1/(2\beta)$ and 
the mass density $\rho$ corresponds to a length scale in the Burgers 
equation. 
In this section, we recast the results of \cite{FraMar99} in the
language of ballistic aggregation of interest here
\footnote{In the present paper, the Brownian motion $P(\tau)$ and the 
parabolic constraints $f_\nu(\tau)$ have the sign opposite to the
corresponding objects in \cite{FraMar99}, namely $-\psi(y)$ and $-s_\nu(y)$.
By invariance of Brownian motion under space reflexion, the functions
$I$ and $J$ are the same in both papers, provided that $M$ and $P$
correspond to $\mu$ and $-\eta$ in the notation of \cite{FraMar99}.}.

\begin{figure}
%\narrowtext
\epsfxsize=11truecm
\hspace{2.75truecm}
\epsfbox{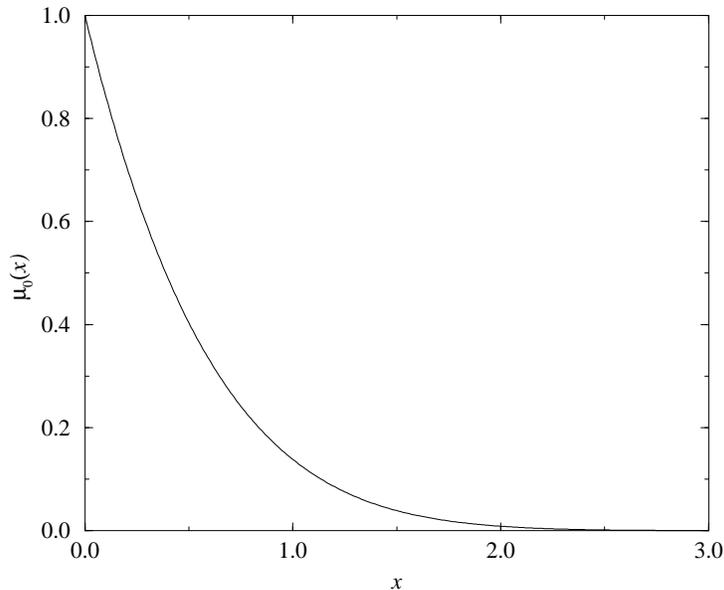}
\caption{The probability density $\mu_0(x)$ to find no aggregates in an 
interval $[0,x]$, for the parameters $t=1$, $\rho=1$ and $\beta=2$ ($a=1$ in
Eq.(\protect\ref{8.46})).}
\label{fig4}
\end{figure}

We consider first the probability density (\ref{7.13}) of finding an 
interval $[0,x)$ void of particles 
\beq
\mu_{0}(x) &=& \int_{-\infty}^\infty dy\, J(y)J(x-y)\non
&=& \sqrt{{\pi\over bx}}
\exp\(-{b^3x^3\over 12}\){1\over (2\pi\,i)^2}\int_{-i\infty}^{+i\infty}
dw_1\int_{-i\infty}^{+i\infty}dw_2\,
{\exp\({bx\over 2}(w_1+w_2)
+{(w_1-w_2)^2\over 4bx}\)\over \Ai(w_1)\Ai(w_2)}
\label{8.46}
\eeq
which is plotted on Fig. \ref{fig4}. 
We have $\lim_{x\to 0}\mu_{0}(x) =1$, and
asymptotically for $x\to \infty$
\be
\mu_{0}(x) =\sqrt{{\pi\over bx}}{\exp\(-{b^3x^3\over 12}-b\omega_1 x\)\over
\left[\Ai'(-\omega_1)\right]^2}\(1+{\cal O}\({1\over x}\)\).
\ee

\begin{figure}
%\narrowtext
\epsfxsize=11truecm
\hspace{2.75truecm}
\epsfbox{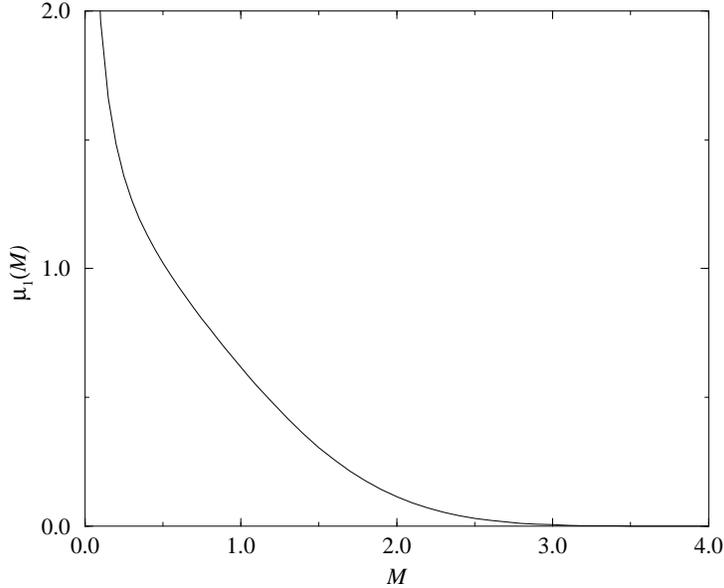}
\caption{The mass density $\mu_1(M)$ for the parameters $t=1$, $\rho=1$
and $\beta=2$ ($a=1$ in Eq.(\protect\ref{pm})).}
\label{fig5}
\end{figure}

The one point probability density (\ref{7.11}) reads
\beq
\mu_1(P,M) &=& I(M,P)J\(-{P\over M}-{M\over 2}\)
J\({P\over M}-{M\over 2}\)\non
&=& 2b^4 {\cal J}\left({P\over M}-{M\over 2}\right)
{\cal I}(M){\cal J}\left(-{P\over M}-{M\over 2}\right)
\label{mu1}
\eeq
where ${\cal I}$ and ${\cal J}$ are given by Eqs.(\ref{cali},\ref{calj}).
Integration over the momentum space yields the mass density 
distribution
\be
\mu_1(M)=2b^3M{\cal I}(M){\cal H}(M)
\label{pm}
\ee
with 
\be
{\cal H}(M)={1\over 2i\pi }\int_{-i\infty}^{+i\infty}
dw{{\rm e}^{-bM w}
\over {\rm Ai}^2(w)}
\label{hhh}
\ee
whose asymptotic behavior reads
\be
{\cal H}(\mu)\sim 1,\;\;\mu\to 0;\quad
{\cal H}(\mu)\sim \sqrt{\pi b^3 \mu^3}\exp\(-{b^3\mu^3\over 12}\)
,\;\;\mu\to \infty.
\label{has}
\ee
The shape of the mass density distribution (\ref{pm}) is plotted on Fig. 
\ref{fig5}. 
Its asymptotic behavior can be computed from (\ref{ias}) and 
(\ref{has}) yielding the formulae
\be
\mu_1(M)=\sqrt{{b^3\over\pi M}}+{\cal O}(M^{1/2}),\quad M\to 0
\label{as1}
\ee
and
\be
\mu_1(M)\sim 2\sqrt{\pi}b^{9/2}M^{5/2}\exp\left(-{b^3M^3\over 12}
-\omega_1 bM\right),\quad M\to\infty.
\label{as2}
\ee
The exact scaling function obtained here is quite different from 
a simple exponential $\exp(-M)$ suggested on the basis of numerical 
simulations in \cite{pom90}.
For example, one notices that small masses ($M\ll t^{2/3}$) are much more
likely to be present in the system than suggested in \cite{pom90}, while 
large masses ($M\gg t^{2/3}$) have a much smaller chance to be present.
Let us remark that the exponential form has been also found in
\cite{tat72} and \cite{jpias92} by solving the dynamical equations in a 
mean-field-like approximation scheme.
In the framework of Burgers turbulence, 
more refined numerical simulations were performed by Kida \cite{kida} where
the small mass behavior of the scaling function compatible with
Eq.(\ref{as1}) has been found. Moreover, our result (\ref{as1}) is 
compatible with the rigorous upper and lower bounds derived in \cite{AveE}.
The large mass behavior Eq.(\ref{as2}) fits into rigorous bounds of the type
$\exp(-CM^3)$ found in \cite{Ave} and \cite{Mar-Pia}.

\begin{figure}
%\narrowtext
\epsfxsize=11truecm
\hspace{2.75truecm}
\epsfbox{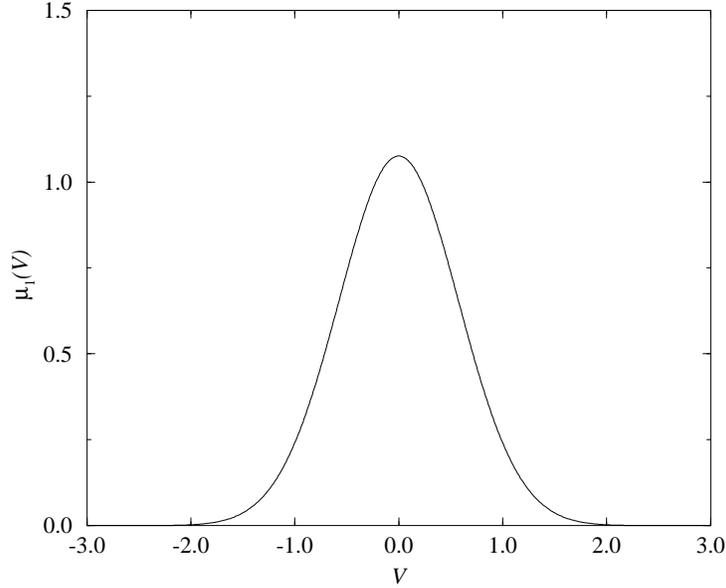}
\caption{The velocity probability density $\mu_1(V)$ for the parameters $t=1$,
$\rho=1$ and $\beta=2$ ($a=1$ in Eq.(\protect\ref{pv})).}
\label{fig6}
\end{figure}

The density of particles of velocity $V=P/M$, given by 
\be
\mu_1(V)=\int_0^\infty dM\,M{\cal I}(M){\cal J}(V-M/2){\cal J}(-V-M/2)
\label{pv}
\ee
is plotted on Fig.\ref{fig6}. 

From (\ref{pv}) and the asymptotic behavior
(\ref{ias}), (\ref{jas}) of $I$ and $J$, one can derive the bound
\be
\mu_1(V)\leq  C|V|\exp\(-{b^3|V|^3\over 3}-b|V|\omega_1\),\quad |V|\to\infty.
\ee
This shows  that the large velocity behavior cannot be Gaussian and thus
invalidates the numerically based hypothesis of Kida \cite{kida}.
 
The density of nearest neighbours (\ref{7.12}) has the form
\beq
\mu_2(1,2)&=&4b^7\theta(X)\delta\(X+Y_1^--Y_2^+\)\non
&&\quad\times 
\exp\(-{b^3\over 3}\((Y_1^-)^3+(Y_2^+)^3\)\){\cal I}(M_1)
{\cal I}(M_2){\cal J}\(-Y_1^+\){\cal J}\(Y_2^-\)
\label{mu2}
\eeq
with $X=X_2-X_1$. Expression (\ref{mu2})
permits to compute the collision frequency between the particles 
defined by
\be
\nu_2(M_1,M_2;t)=\int_{-\infty}^\infty dP_1
\int_{-\infty}^\infty dP_2\, \left|{P_1\over M_1}-{P_2\over M_2}\right|
\mu_2(0,P_1,M_1,0+,P_2,M_2;t).
\ee
First, we notice that relation (\ref{7.12b}) implies the scaling
\be
\nu_2(M_1,M_2;t)={\rho^2\over (\rho t)^3}\nu_2(M_1^\prime ,M_2^\prime )
\ee
with
\be
\nu_2(M_1,M_2)=\int_{-\infty}^\infty dP_1
\int_{-\infty}^\infty dP_2\,\left|{P_1\over M_1}-{P_2\over M_2}\right|
\mu_2(1,2)
\ee
This last integral, upon inserting formula (\ref{mu2}),  gives
\be
\nu_2(M_1,M_2)=2b^6M_1M_2(M_1+M_2){\cal I}(M_1){\cal I}(M_2)
{\cal H}(M_1+M_2)
\ee
with ${\cal H}$ given by Eq.(\ref{hhh}).
Clearly, the collision frequency does not factorize into a product of a
function of $M_1$ and a function of $M_2$. This fact invalidates the weak 
mean-field hypothesis whose consequences where analysed in \cite{jpias92}.   

\vspace{5mm}

{\bf The two-point distribution function}

\vspace{5mm}

The scaling form $\rho_2(1,2)=\rho_1\rho_2(2|1)$ of the two-point reduced 
number density of aggregates  enables to study the long distance
correlations in the system. The calculation requires the summation over
all possible configurations of aggregated masses 
in between the particles $1$ and $2$ (formula (\ref{40})). 
In principle, this summation can be  be performed as follows. 
Because of translation invariance, the conditional probability $\mu (2|1)$ 
\be
\mu (X_{2},P_{2},M_{2}|X_{1},P_{1},M_{1})= 
(P_{1},M_{1}|T(X_{2}-X_{1})|P_{2},M_{2})
\label{co2}
\ee
can be considered as the kernel of an integral operator $T(X)$ acting
in the  mass and momentum space. Then the series (\ref{40}) 
can be summed up by applying the Laplace transformation. Setting
\beq
\tilde{T}(s)&=&\int_{0}^{\infty}dXe^{-Xs}T(X)\non
\tilde{\rho}_{2}(s,P_{1},M_{1},P_{2},M_{2})
&=&\int_{0}^{\infty}dXe^{-Xs}\rho_{2}(0,P_{1},M_{1},X,P_{2},M_{2})
\label{co3}
\eeq
one finds by the convolution theorem
\beq
\tilde{\rho}_{2}(s,P_{1},M_{1},P_{2},M_{2})&=&\rho_{1}(P_{1},M_{1})\sum_{r=1}^{\infty}(P_{1},M_{1}|\tilde{T}^{r}(s)
|P_{2},M_{2})\non
&=&\rho_{1}(P_{1},M_{1})(P_{1},M_{1}|\tilde{T}(s)(I-\tilde{T}(s))^{-1}|P_{2},M_{2})
\label{co4}
\eeq
Inverting the operator $I-\tilde{T}(s)$ and then performing the inverse 
Laplace transformation are not easy operations here. 
So, we shall proceed in a different way.
In the context of the Burgers equation, $\rho_{2}(1,2)$ corresponds to the
joint density to find two shocks at distance $|X_{2}-X_{1}|$. 
This density, in the Brownian interpretation,
is the measure of paths that have contact points with two parabolas: one
centered at $X_{1}$ another at $X_{2}$. Summing over all such paths
amounts to sum over intermediate sequences of neighbouring aggregates in
the particle language. We refer to \cite{FraMar99} for details.  One finds 
the following form of the two-point density 
\beq
\rho_{2}(1,2) &=& J\(-Y_1^+\)I(M_1,P_1)\left[\delta\(X+Y_1^--Y_2^+\)\right.\non
&&\quad\left. +\theta\(X+Y_1^--Y_2^+\)H\(X,Y_1^-,Y_2^+\)\right]
I(M_2,P_2)J\(Y_2^-\)
\eeq
The $\delta-$function term corresponds to the case when $1$ and $2$
are the nearest neighbours, whereas the function $H\(X,Y_1,Y_2\)$ embodies 
precisely  the sum over all configurations of intermediate particles. The 
function $H$ will not be reproduced here. The important fact is that its 
large distance asymptotic behaviour can be calculated leading to the folowing 
cluster property of the two-point function
\cite{FraMar99}
\beq
\rho_{2}(M_1,P_1,&& M_2,P_2,X)
-\rho_{1}(M_{1},P_{1})\rho_{1}(P_{2},M_{2})
\sim -b^{11/2}{32\sqrt{\pi}\over X^{5/2}}
\exp\(-\frac{b^3X^{3}}{12}-b\omega_{1}X\)\non
&& \times \exp\(-b\omega_{1}(Y_1^--Y_2^+)\)
{\cal J}\(-Y_1^+\){\cal I}(M_1){\cal I}(M_2){\cal J}\(Y_2^-\)
,\quad X\to\infty.
\label{c0a}
\eeq
We see that although the particles in the initial state
are not correlated and the motion between collisions
is free, the aggregation process induces dynamic correlations in the
course of time. However, these correlations have a very short range since 
they stay dominated by the rapidly decaying cubic exponential 
factor $\exp(-b^3X^3/12)$.

\section{Rigorous solution of the dynamic hierarchy}\label{sec10}

We have shown in Section \ref{sec4} 
how the infinite  hierarchy (\ref{23}) describing
the aggregation process could be reduced to a system of two coupled equations
(\ref{32}), (\ref{33}), satisfied by the one-particle density $\mu_{1}(1;t)$, 
and the conditional probability density $\mu (2|1;t)$. It is the right moment
now to show that the constructed densities (\ref{7.11}), (\ref{7.12}) provide 
a rigorous solution to the aggregation dynamics. 
 
In (\ref{7.11}) and (\ref{7.12}) the densities $\mu_{1}(1;t)$ and 
$\mu (2|1;t)$ are expressed in terms of the probability weights $I(M,P)$ and
$J(Y)$ (to recover the time variable $t$ the 
scaling relations (\ref{7.100}) have to be used). In order to evaluate the 
rate of change of the state of the aggregating gas it is thus sufficient 
to calculate the time derivative of functions $I(M,P;t)$, $J(-Y^{+};t)$ 
and $J(Y^{-};t)$, where $Y^{\pm }=tP/M \pm M/2\rho $. 
A straightforward calculation can be performed starting from the defining 
formulae (\ref{6.7}), (\ref{6.8}), whose explicit form can be found
in \cite{Mar-Pia}. Technically it is simple but lengthy, so we shall not
reproduce it here. Taking then the continuum limit one finds the following 
system of equations
\beq
\frac{\partial}{\partial t}I(M,P;t) &=& \int dM_{1}\int dP_{1} \int dM_{2}
\int dP_{2} \left( \frac{P_{1}}
{M_{1}}-\frac{P_{2}}{M_{2}}\right)\delta (M-M_{1}-M_{2}) 
\delta (P-P_{1}-P_{2})\non
&&\quad \times \delta \left[ 
\left( \frac{P_{1}}{M_{1}}-\frac{P_{2}}{M_{2}}\right)  
-\frac{M}{2\rho}\right] I(M_{1},P_{1};t)I(M_{2},P_{2};t)  \label{10.1}
\eeq
\beq
\frac{\partial}{\partial t}J\left( \frac{P}{M}t-\frac{M}{2\rho };t\right) &=& 
-\int dM_{1}\int dP_{1} \left( \frac{P}{M}-\frac{P_{1}}{M_{1}}\right)
\delta \left[ \left( \frac{P}{M}-\frac{P_{1}}{M_{1}}
\right)t -\frac{M+M_{1}}{2\rho } \right] \non
&& \quad 
\times I(P_{1},M_{1};t)J\left(\frac{P_{1}}{M_{1}}t-\frac{M_{1}}{2\rho };t
\right) \label{10.2} 
\eeq

A somehow laborious but straightforward  analysis permits to check that 
the above system of equations, when combined with the relations (\ref{7.11}), 
(\ref{7.12}), implies the fundamental dynamic equations (\ref{32}), (\ref{33}).
As one could expect from the interpretation of functions $I$ and $J$ given in
Section \ref{sec7}, the time derivative of $I(M,P;t)$ is positive and produces 
gain terms in the kinetic equation (\ref{32}), whereas the negative time 
derivatives of factors $J(Y^{+};t)$ and $J(Y^{-};t)$ generate loss terms. 

\section{concluding comments}

The hierarchy equations (\ref{23}) describe the aggregation dynamics in a
one-dimensional gas for arbitrary initial conditions. We considered
here in detail the case where statistical 
correlations  existed only between the states of the nearest neighbours. Under
the precise condition (\ref{00}), the
evolution preserved this property, and the state of the system remained 
entirely characterized by the number density of aggregates and by the 
conditional distribution of nearest neighbours. This permitted to reduce
the hierarchy to two coupled equations.

Rather than solving the obtained system of equations we determined the 
relevant distributions on the basis of their definitions, and only a 
posteriori we could verify that they satisfied the dynamical laws. 
So, finding a systematic way 
of solving the system (\ref{32}), (\ref{33}) remains an open problem.

When discussing the continuum limit we started from a very simple discrete
initial condition, where no correlations were present at all. In the course
of time, the system turned out to build up correlations by its internal 
dynamics, but only between the pairs of nearest neighbours. It is thus 
tempting to conjecture that the factorized form (\ref{31}) of the 
many particle densities represents the asymptotic long time structure of the
system for  general initial conditions, as only the nearest
neighbours get correlated by the process of aggregation. Let us recall that
the correlations are rapidly decaying  with the distance (see (\ref{c0a}).

A comment concerning the continuum limit seems also quite important. It has 
been noted at the end of Section \ref{sec7} that introduction of properly scaled
variables permitted to interpret the continuum limit as a long time limit. 
One can thus expect that the dynamical properties derived here are quite
universal, and do not depend on our particularly simple choice of the initial
distribution. In particular, it would be interesting to clarify how large is
the class of initial states which leads for long times to the mass density
distribution plotted in Fig.(\ref{fig5}).  

Let us finally notice, that the dynamical scaling of 
the mass distribution predicted on the basis of intuitive arguments in 
\cite{pom90} agrees with the rigorous scaling relations (\ref{6.9}).  
The prediction of the Brownian motion exponent $\alpha = 2/3$ in
\cite{pom90} seems independent of the initial condition, whereas we
derived it starting from a particular state. It would be thus
interesting to clarify to what extent this type of scaling is universal.
Indeed,  our analysis showed that the aggregation dynamics was in principle
compatible with other values of $\alpha$, not necessarily equal to $2/3$.

\acknowledgements

J. Piasecki acknowledges the hospitality at the Institute of Theoretical
Physics of the Ecole Polytechnique F\'ed\'erale de Lausanne, and the 
financial support by the KBN (Committee for Scientific Research, Poland),
grant 2 P03B 127 16.

\appendix
\section{}
\label{app1}

The moments of the distribution $G_{k}$ (\ref{7.20}) can be obtained from its
generating function $\gamma(s,\xi)$
\be
\sum_{k=1}^\infty s^k \left\langle 
q^j\right\rangle_k
%=\sum_{k=1}^\infty s^k \int_0^\infty dq\, 
% q^r G_k(q)
=\left.(-i)^j{\partial^j \gamma(s,\xi)\over \partial \xi^j}\right|_{\xi=0}
\label{A1}
\ee
with
\be
\gamma(s,\xi)=1+\sum_{k=1}^\infty s^k \int_0^\infty dq\,
{\rm e}^{i\xi q}G_k(q)
\label{A2}
\ee
This generating function can be computed by applying the Sparre-Andersen 
theorem to the process with independent increments governed by the Gaussian
distribution $(\pi)^{-1/2}\exp(-q^{2})$ (section XVIII.3 of \cite{Fel})
\be
\gamma(s,\xi)=\exp\left[\sum_{k=1}^\infty {s^k\over k}
\int_0^\infty dq\,{\rm e}^{i\xi q}
{{\rm e}^{-q^2/k}\over \sqrt{\pi k}}\right].
\label{A3}
\ee
Using (\ref{A3}) we find
\be
\sum_{k=1}^\infty s^k \left\langle 
1\right\rangle_k=
\gamma(s,0)=\exp\left(\sum_{k=1}^\infty {s^k\over 2k}\right)
={1\over\sqrt{1-s}}
\label{A5}
\ee
and thus 
\be
\left\langle  1\right\rangle_k={\Gamma(k+1/2)
\over\sqrt{\pi} \Gamma(k+1)}\sim {1\over\sqrt{\pi k}},\quad k\to\infty.
\ee
For the first and second moments we find
\be
\sum_{k=1}^\infty s^k \left\langle  q\right\rangle_k=
-i\left.{\partial \gamma(s,\xi)\over \partial \xi}\right|_{\xi=0}=
{1\over\sqrt{1-s}}\sum_{k=1}^\infty {s^k\over 2\sqrt{\pi k}}
\label{A6}
\ee
and 
\be
\sum_{k=1}^\infty s^k \left\langle  q^2\right\rangle_k=
-\left.{\partial^2 \gamma(s,\xi)\over \partial \xi^2}\right|_{\xi=0}=
{1\over 4 \sqrt{1-s}}\left[{1\over 1-s}+
\left(\sum_{k=1}^\infty {s^k\over \sqrt{\pi k}}\right)^2\right].
\label{A7}
\ee
To proceed, we need the following Tauberian theorem (Theorem 5, 
section XIII.5 in \cite{Fel}):

Let $c_{k}\geq 0$ be a monotonic sequence and suppose that the 
series $\sum_{k=1}^{\infty}c_{k}s^{k}$ converges for $0\leq s <1$. Then, 
if $L$ varies slowly at infinity 
\footnote{$L$ varies slowly at infinity if $\lim_{t\to\infty}L(tx)/L(t)=1$ 
for all $x$, a condition which is  
verified in the subsequent applications where $L$ is a constant.} 
and $\alpha\geq 0$, each of the two relations
\be
\sum_{k=1}^{\infty}c_{k}s^{k}\sim \frac{1}{(1-s)^{\alpha}}
L\(\frac{1}{1-s}\),\quad s\to 1
\label{A8}
\ee
\be
c_{k}\sim \frac{1}{\Gamma(\alpha)}k^{\alpha-1}L(k),\quad k\to\infty
\label{A9}
\ee
implies the other.

From (\ref{A6}), we have
\be
\sum_{k=1}^\infty s^k \left\langle  q\right\rangle_k
\sim {1\over 2(1-s)},\quad s\to 1
\ee
and thus
\be
\left\langle  q\right\rangle_k\sim {1\over 2},\quad  k\to \infty.
\ee
For the second moment we find
\be
\sum_{k=1}^\infty s^k \left\langle 
q^2\right\rangle_k\sim {1\over 2(1-s)^{3/2}},\quad s\to 1
\ee
leading to
\be 
\left\langle  q^2\right\rangle_k
\sim {\sqrt{k}\over \sqrt{\pi}},\quad  k\to \infty.
\ee

To control the remainder in (\ref{7.23}) by a limited Taylor expansion up 
to second order in $q$ we use the fact that $G_{k}(q)$ is non negative and 
that the second $q$-derivative of ${\rm e}^{-\nu\sqrt{2m\beta}q}
K_{m,\nu}(M_0,\nu M_0+\sqrt{{2m\over\beta}}q,M,P)$ has a 
$q$-integrable 
bound for $M>0$ uniform when $m$ and $M_{0}$ are in a neighbourhood of zero
(irrespective of the sign of $\nu$). 
This is indeed the case because one can check that 
$K_{\nu}(M_{0},P_{0},M, P)$ as well as its first and second 
$P_{0}$-derivative obey  Gaussian bounds of the form
$C_{1}\exp(-C_{2}P_{0}^{2})$ with $C_{1}$ and $C_{2}$ independent 
of $m$ and $M_{0}, \;M_{0}<M$ \\[1cm]

\end{document}